\documentclass[twocolumn]{revtex4-2} 

\usepackage{graphicx} 
\usepackage{amsmath}
\usepackage{amssymb}

\begin{document}

\title{An axiomatic limitation on the deterministic scope required for superdeterminism and its consequentially greater likelihood}
\author{Cameron Shackell}
\affiliation{Queensland University of Technology}

\begin{abstract}
By positing a universe where all events are determined by initial conditions, superdeterminism as conceded by Bell frames correlations observed in quantum measurements as the consequence of an inherently predetermined cosmic order that shapes even our experimental choices. I use an axiomatic formulation of superdeterminism to demonstrate that Bell overstated the scope of determinism required. Assuming only the existence of a universe containing observers, I show that determinism in just the observer scope is sufficient. I then discuss how this sufficiency increases the theory's plausibility and suggest a path to its integration with results from other disciplines.
\end{abstract}

\maketitle

\section{Introduction}

Quantum anomalies present a profound challenge to our understanding of the universe. Wave-particle duality\cite{Einstein1905}\cite{Dirac1930}, quantum entanglement \cite{Einstein1935}\cite{Bell1964}, and retrocausality \cite{WheelerFeynman1945}\cite{Cramer1986} fly in the face of intuitive notions of space and time.

Superdeterminism, conceded by Bell \cite{davies1993ghost} as a complete but implausible explanation, offers an alternative to theories that rely on nonlocal mechanisms famously described by Einstein as ``spooky action at a distance'' (``spukhafte Fernwirkung'') \cite{EinsteinBornLetters}. By positing a universe where all events are determined by initial conditions, superdeterminism frames correlations observed in quantum measurements as the consequence of an inherently predetermined cosmic order---one that shapes even our experimental selections and behavior.

An unpalatable aspect of superdeterminism for many is that it invites a reevaluation of statistical independence, jeopardizing the validity of science and notions of free will. Furthermore, perhaps because the idea is so ancient, final, and unnuanced, superdeterminism is often seen as a cheat's or conspirator's solution, akin perhaps to x=0 in certain differential equations.

In this paper I take an axiomatic perspective on the scope required for superdeterminism to hold. Assuming only the existence of states of a universe containing observers, I show that the universal ``super'' scope Bell proposed is an unnecessary hyperbole and that a revised formulation lends subtlety, increases credibility, and opens a path to integration with results from other disciplines.

\section{An axiomatic formulation of superdeterminism}

An accessible recent discussion of superdeterminism outlining common features of superdeterministic quantum theories is provided by Hossenfelder and Palmer \cite{HossenfelderPalmer}. In rough alignment, the following model formally sketches how superdeterminism can be enlisted as a general explanation for quantum anomalies.

\subsection{Axioms of superdeterminism}

\begin{enumerate}
  \item \textbf{Axiom of Predetermined Outcomes:}
  \begin{itemize}
    \item Let \( \Psi(t, X) \) represent the state of the universe at time \( t \) with initial condition \( X \).
    \item The evolution of \( \Psi \) is governed by a deterministic function \( F \), such that \( \Psi(t, X) = F(t, \Psi(0, X)) \), where \( \Psi(0, X) \) is the state of the universe at the initial time (t=0).
  \end{itemize}

  \item \textbf{Axiom of Mapping to Initial Conditions:}
  \begin{itemize}
    \item The initial condition \( X \) fully determines \( \Psi \) for all future times: \( \forall t > 0, \, \Psi(t) = G(X) \), where \( G \) is a deterministic mapping from initial conditions to future states.
  \end{itemize}
\end{enumerate}

\subsection{Corollaries explaining quantum anomalies}

  \begin{enumerate}
      \item \textbf{Measurement Correlation:}
      \begin{itemize}
      \item Let \( M(t) \) represent a measurement at time \( t \), and \( O(t) \) the outcome.
      \item Both \( M(t) \) and \( O(t) \) are functions of \( \Psi(t, X) \), indicating correlation: \( M(t), O(t) \leftarrow H(\Psi(t, X)) \), where \( H \) maps the state of the universe to measurement settings and outcomes.
      
      \end{itemize}
      
      \item \textbf{Irrelevance of Observer Choice:}
      \begin{itemize}
      \item The observer's choice \( C(t) \) is also some function of \( \Psi \): \( C(t) = K(\Psi(t, X)) \), indicating that it is predetermined and correlated with outcomes.
      \end{itemize}

\end{enumerate}
      
\section{Universal scope: the ``super'' of Bell's superdeterminism}

The ``super'' in superdeterminism, which can be regarded as a pleonasm, was prepended by Bell to emphasize the ``absolute determinism'' he thought required\cite{davies1993ghost}. It underscores his position that, to have explanatory power for phenomena like entanglement, all events need to have been determined at the very beginning of the universe, and to have evolved in a deterministic way through all time and space to converge at any measurement event. The prefix ``super'' excludes out of hand the possibility of limited determinism, insisting that violation of statistical independence requires observations everywhere to be correlated.

Bell was perhaps more rhetorically savvy than rigorous in this formulation. Superdeterminism's extreme ``universal scope'' is exactly what dissuades many physicists from it, not least because, if all measurement is ineluctably correlated, it invalidates the scientific method - a tool that seems nonetheless to have proven fruitful over the last centuries and underpins most current research.

Precisely because Bell took pains to be so emphatic, however, is good reason to examine the idea of the universal ``super'' scope more closely.

\section{The universal scope and the observer scope}

It is self-evident that the universal scope contains the observer. In fact, this is implicit in the corollary of superdeterminism used to explain quantum anomalies (included in the sketched model above as ``Irrelevance of Observer Choice'') . We can formalize this as:

\begin{description}

  \item[\textbf{Corollary 3}] Observer Inclusion\\\textit{The state of the universe includes the states of all observers.}

\end{description}

This corollary  allows a division of the universal scope into matter relevant to observers (often called the ``observer substrate'') and matter not relevant to observers. For clarity, these are  referred to hereafter as the \textit{observer scope} and the \textit{non-observer scope} which, combined, constitute the \textit{universal scope}.

\section{Equivalence of the observer and universal scopes}

Despite being crucial to superdeterminism explaining quantum anomalies, there is a clear relativistic problem in separating the observer and non-observer scopes. How can anything outside the observer scope ever be relevant to observation? Even taking an ontologically minimalist position and making no claims or assumptions about the extent, character, or properties of observation, the following must nonetheless be true:

\begin{description}

    \item[\textbf{Axiom 3}] Equivalence of observer and universal scopes \\
    \textit{Any observation of the universe is contained, by definition, in the states of observers. Therefore, the universal scope and the observer scope cannot be distinguished by observers.}
\end{description}

The correctness of this corollary can be confirmed counterfactually: one cannot observe the universe without being an observer; to do so would be to pretend that observation can occur beyond the observer scope.

This equivalence leads to the following important corollary:

\begin{description}
  \item[\textbf{Corollary 5}] Observer scope limitation \\ \textit{As the observer and universal scopes are indistinguishable, superdeterminism only requires determination of the observer scope.}
\end{description}

The question of what, if anything, remains in the non-observer scope depends upon what ontological commitments are introduced about observation. To canvas just three possibilities:

\begin{enumerate}
    \item The non-observer scope is empty
    \item The non-observer scope is non-empty and deterministic
    \item The non-observer scope is non-empty and non-deterministic
\end{enumerate}

None of these, however, are relevant to superdeterminism explaining quantum anomalies because they lie, by definition, outside the observer scope.

\section{Revised axiomatic formulation of superdeterminism}

Under the equivalence of the the universal scope and the observer scope derived above, the axiomatic sketch of superdeterminism can be rewritten as follows.

\subsection{Axioms of superdeterminism limited to the observer scope}

\begin{enumerate}

\item \textbf{Axiom of equivalence of the universal and observer scopes:}
\begin{itemize}
    \item At any given time \( t \), observation of the universal state is completely contained in the states of all observers.
    \end{itemize}
    
  \item \textbf{Axiom of Predetermined Outcomes:}
  \begin{itemize}
    \item Let \( \Psi(t, X) \) represent the state of all observers at time \( t \) with initial condition \( X \).
    \item The evolution of \( \Psi \) is governed by a deterministic function \( F \), such that \( \Psi(t, X) = F(t, \Psi(0, X)) \), where \( \Psi(0, X) \) is the state of all observers at the initial time (t=0).
  \end{itemize}

  \item \textbf{Axiom of Initial Conditions:}
  \begin{itemize}
    \item The initial condition \( X \) fully determines \( \Psi \) for all future times: \( \forall t > 0, \, \Psi(t) = G(X) \), where \( G \) is a deterministic mapping from initial conditions to future states.
  \end{itemize}

 \end{enumerate}

\subsection{Corollaries explaining quantum anomalies}

\begin{enumerate}
      \item \textbf{Measurement Correlation:}
      \begin{itemize}
      \item Let \( M(t) \) represent a measurement at time \( t \), and \( O(t) \) the outcome.
      \item Both \( M(t) \) and \( O(t) \) are functions of \( \Psi(t, X) \), indicating correlation: \( M(t), O(t) \leftarrow H(\Psi(t, X)) \), where \( H \) maps the state of observers to measurement settings and outcomes.
      
      \end{itemize}
      
      \item \textbf{Observer Choice:}
      \begin{itemize}
      \item The observer's choice \( C(t) \), being by Axiom 1 contained in the state of the observer, is precisely the function \( \Psi \), indicating that it is predetermined.
  \end{itemize}
\end{enumerate}

\section{Implications for the likelihood of superdeterminism}

Simply by Occam's razor, the likelihood of superdeterminism is increased by showing that the conditions for it are satisfied by restriction to a smaller scope. Depending on one's view of the non-observer scope, that increase may be small (in the case that the non-observer scope is small) or quite appreciable (in the case that the non-observer scope is comparatively large).  In all cases, Bell's formulation which requires determinism in the entire universe reduces to determinism merely in matter constituting observers.

\section{Is determinism in the observer scope plausible or useful?}
Two questions immediately follow from the revised formulation of superdeterminism above:

\begin{enumerate}
    \item Does any evidence support determinism in the observer scope?
\end{enumerate}

Essentially, this question asks: is there any reason to believe that the observer substrate evolves in a deterministic way? Of course, no simple answer will suffice, but it is perhaps not facile to point out that the theory of evolution and the study of biological systems suggest precisely this\cite{Darwin1859}\cite{Dobzhansky1937}\cite{WatsonCrick1953}. Moreover, various perspectives on neurology\cite{Penfield1975} and linguistics\cite{Whorf1956} might be seen to align.

\begin{enumerate}
    \item[2.] Does assuming determinism in the observer scope reveal avenues for better understanding of quantum phenomena?
\end{enumerate}
This question is also of enormous complexity. At the very least, observer rather than universal determinism allows for novel restatement of some core concepts in quantum foundations. For example, the violation of statistic independence might be interestingly restated as an outcome of biological speciation as follows:

\begin{itemize}
    \item[] Under determinism of the observer substrate, statistical independence may be violated when all observers have, at some remote point, a common phylogeny, ontogeny, or environmental influence occasioning a correlated tendency to choose experiments, differentiate objects, and make measurements.
\end{itemize}

In terms of explaining quantum anomalies, observer determinism urges attention to the relation of the observer scope to the hypothetical non-observer scope via careful \textit{representation}. For example, the observer scope might be posited as dynamic---expanding to include various elements of the non-observer scope as required. This material expansion and contraction, however, is not naively isomorphic with representation. When an observer measures the apparent width of a galaxy, for example, they do not draw the entire galaxy into the observer scope. The same may perhaps be said of quantum particles.

\section{Conclusion}

If one accepts that observer states and thus observation emerge from matter, one must accept axiomatically that superdeterminism only requires determinism of that observer-relevant matter. To argue otherwise is to demand a transcendental treatment of the observer substrate.

By narrowing the scope required for superdeterminism from the universal to the observer scope, a significant improvement is made in its plausibility. The imposing task of divining a deterministic basis for everything in the universe becomes a much more immediate and intuitively rich invitation to explore the ramifications of material determination of observation, for which there is evidence from, among others, the study of biological systems and neurology.


\bibliographystyle{naturemag}
\bibliography{refs.bib}

\end{document}